\documentclass{article}

\usepackage{arxiv}

\usepackage[utf8]{inputenc} 
\usepackage[T1]{fontenc}    
\usepackage{hyperref}       
\usepackage{url}            
\usepackage{booktabs}       
\usepackage{amsfonts}       
\usepackage{nicefrac}       
\usepackage{microtype}      
\usepackage{lipsum}		
\usepackage{graphicx,amssymb,amsmath,color}
\usepackage{doi}
\usepackage{array}
\newcolumntype{P}[1]{>{\centering\arraybackslash}p{#1}}

\usepackage[table,x11names]{xcolor}
\usepackage{dcolumn}
\usepackage{setspace}
\usepackage{rotating}
\usepackage{afterpage}
\usepackage[numbers,sort&compress]{natbib}

\newcolumntype{.}{D{.}{.}{-1}}

\usepackage{authblk}
\newcommand{\fig}{Fig.~}
\newcommand{\figs}{Figs.~}
\newcommand{\eqn}{Eq.~}
\newcommand{\eqns}{Eqs.~}
\newcommand{\tbl}{Table~}

\newcommand{\sectn}{\S~}

\definecolor{blue}{rgb}{0, 0.5, 0.5}
\definecolor{blue2}{rgb}{0.1216, 0.4667, 0.7059}
\definecolor{red}{rgb}{0.8500, 0.1250, 0.0480} 
\definecolor{red2}{rgb}{0.8392, 0.1529, 0.1569} 
\definecolor{orange2}{rgb}{1.0, 0.498, 0.0549} 
\definecolor{yellow2}{rgb}{0.9290, 0.6940, 0.1250}
\definecolor{purple}{rgb}{0.4940, 0.1840, 0.5560}
\definecolor{purple2}{rgb}{0.5804, 0.4039, 0.7412}
\definecolor{green}{rgb}{0.4660, 0.6740, 0.1880}
\definecolor{green2}{rgb}{0.1725, 0.6275, 0.1725}
\definecolor{ltblue2}{rgb}{0.0902, 0.7451, 0.8118}
\definecolor{dkred2}{rgb}{0.6350, 0.0780, 0.1840}
\definecolor{gray2}{rgb}{0.22, 0.22, 0.3}
\definecolor{gray3}{rgb}{0.5, 0.5, 0.5}

\title{Machine-Learned Closure of URANS for stably stratified turbulence: Connecting physical timescales \& data hyperparameters of deep time-series models}

\author[a,$\dagger$]{Muralikrishnan Gopalakrishnan Meena}
\author[b]{Demetri Liousas}
\author[b,1]{Andrew D. Simin}
\author[a]{Aditya Kashi}
\author[a]{Wesley H. Brewer}
\author[c]{James J. Riley}
\author[b]{Stephen M. de Bruyn Kops}

\affil[a]{National Center for Computational Sciences, Oak Ridge National Laboratory, Oak Ridge, TN 37831, USA}
\affil[b]{Department of Mechanical and Industrial Engineering, University of Massachusetts Amherst, Amherst, MA 01003, USA}
\affil[c]{Department of Mechanical Engineering, University of Washington, Seattle, WA 98105, USA}
\affil[1]{Now at DoD HPCMP PET/GDIT, Vicksburg, MS 38180, USA}
\affil[$\dagger$]{To whom correspondence should be addressed: \href{mailto:gopalakrishm@ornl.gov}{gopalakrishm@ornl.gov}}

\date{}

\renewcommand{\headeright}{~}
\renewcommand{\undertitle}{~}
\renewcommand{\shorttitle}{Interpretable Deep Time-Series ML Closure of URANS for SST}

\begin{document}

\maketitle

\begin{abstract}
We develop time-series machine learning (ML) methods for closure modeling of the Unsteady Reynolds Averaged Navier Stokes (URANS) equations applied to stably stratified turbulence (SST). SST is strongly affected by fine balances between forces and becomes more anisotropic in time for decaying cases. Moreover, there is a limited understanding of the physical phenomena described by some of the terms in the URANS equations. Rather than attempting to model each term separately, it is attractive to explore the capability of machine learning to model groups of terms, i.e., to directly model the force balances. We consider decaying SST which are homogeneous and stably stratified by a uniform density gradient, enabling dimensionality reduction. We consider two time-series ML models: Long Short-Term Memory (LSTM) and Neural Ordinary Differential Equation (NODE). Both models perform accurately and are numerically stable in \textit{a posteriori} tests. Furthermore, we explore the data requirements of the ML models by extracting physically relevant timescales of the complex system. We find that the ratio of the timescales of the minimum information required by the ML models to accurately capture the dynamics of the SST corresponds to the Reynolds number of the flow. The current framework provides the backbone to explore the capability of such models to capture the dynamics of higher-dimensional complex SST flows.

\textbf{Notice:} This manuscript has been authored by UT-Battelle, LLC, under contract DE-AC05-00OR22725 with the US Department of Energy (DOE). The US government retains and the publisher, by accepting the article for publication, acknowledges that the US government retains a nonexclusive, paid-up, irrevocable, worldwide license to publish or reproduce the published form of this manuscript, or allow others to do so, for US government purposes. DOE will provide public access to these results of federally sponsored research in accordance with the DOE Public Access Plan (\href{http://energy.gov/downloads/doe-public-access-plan}{http://energy.gov/downloads/doe-public-access-plan}).
\end{abstract}


\section{Introduction}
\label{sec:intro}
Stably stratified turbulence (SST) is a model for understanding fluid flows that involve an ambient stable density stratification in the oceans due to temperature or salt gradients.  Studying SST has important implications for climate modeling, pollution mitigation, and deep sea mining.  In general the dynamics of SST reflect a competition between many forces including inertial, viscous, buoyancy, Coriolis, and shearing.  Even in the simplest case, though, SST can be parameterized by a minimum of two dimensionless groups defined by ratios of the inertial, buoyancy, and viscous forces.  These ratios can be chosen in a variety of ways, and we use Reynolds and Froude numbers defined later.  Consistent with the parameter space of SST being, at a minimum, two dimensional, the flows are highly intermittent and anisotropic at large scales.  \citet{riley12} provide an overview of SST, \citet{debk19} review many theoretical, laboratory, and numerical studies in SST, and \citet{mater14} present a framework for parameterizing a three parameter variant of SST that includes mean shear in order to describe the flow dynamics in terms of where the flow resides in parameter space.

The extreme scales required to capture the evolution of such flows in nature hinder computational experiments even with the latest high-performance computing platforms \cite{norman2021unprecedented} and thus, the atmospheric and oceanographic communities have relied on parametrizations or surrogate models to capture the effect of the small scale flow features in SST. Turbulence closure models in Unsteady Reynolds Averaged Navier Stokes (URANS) simulations and large-eddy simulations (LES) of the atmosphere, ocean, and their sub-systems are broad classes among many such turbulence models. Particularly, URANS second-moment closure (SMC) based modeling has been prevalent in simulations of stratified shear layers and wakes \cite{jain2022second,jain2022study}. Such models are limited by the appropriate choice of the parametrization, interpretation of the parametrizations, and generalizability with different flow regimes and types. Data-driven techniques are attractive to overcome such challenges. Over the past decade, there is a history of using machine learning (ML) methods to tackle fluid flow problems \cite{brunton2020machine}, and specifically, to assist in the augmentation of turbulence models \cite{duraisamy_review_2019}. The atmospheric and oceanographic communities have been using ML to formulate surrogate models to capture the effects of sub-grid scale structures. Such efforts have been focused on turbulence closures for LES and global climate models \cite{rasp2018deep,yuval2020stable,krasnopolsky2013using,gentine2018could,brenowitz2019spatially,watt2021correcting,bretherton2022correcting,cheng2022deep}. There have been various efforts to use data-driven models in modeling other applications of stratified turbulent flows as well, such as shear layers \cite{salehipour2019deep}, stratified wakes \cite{huang2023linear}, engineering applications \cite{huang2019wall,oda2022study,xu2024data}, optical turbulence \cite{bolbasova2021application}, and mixing in SST \cite{couchman2021data,lewin2023data,lefauve2023data,petropoulos2023prandtl,lefauve2023routes}.

Developing predictive capability for SST is the first motivation for this study.  A second is to apply machine learning (ML) to the difficult problem of closing the URANS equations or the related approach of LES.  In URANS, differential equations for averaged quantities are solved numerically with models for terms involving fluctuations about the averages.  In LES, equations are solved for quantities resolved on the numerical grid and subgrid-scale quantities are modeled.  With URANS, even in configurations involving just two types of forces, e.g., inertial and viscous, all but the simplest closure models typically involve differences between model terms.  SST flows involve more than two forces, so there are more terms to be modeled in either URANS or LES, and there is limited understanding of the physical phenomena described by some of these terms. Rather than attempting to model each term separately, as is done in SMC modelling, it is attractive to see how well ML can be used to model groups of terms, that is, to directly model the force balances.

Furthermore, the inherent nonlinear dynamics of the SST, even in a reduced-order sense, pose the challenge of finding interpretable and generalized surrogates. Thus, the third and final motivation of this study is to formulate interpretable ML models that accurately predict force balances. Specifically, we will focus on interpreting the data requirements of time-series-based ML models for modeling a reduced dimensional representation of canonical SST.

Advances in ML have shown the capability of ML models to implicitly learn the underlying dynamics of a time-dependent physical system \cite{connor1994recurrent,yu2019review,che2018recurrent,hewamalage2021recurrent}. Recurrent neural networks, such as Long short-term memory (LSTM) \cite{hochreiter1997long,hua2019deep,lindemann2021survey} and gated recurrent units (GRUs) \cite{chung2014empirical,weerakody2021review}, have been shown to model and predict time-series data effectively.
Thus, time-series ML models could provide a computationally inexpensive way to simulate complex time evolving force balance of the SST URANS equations. The task of interpreting such models still remains a challenge \cite{arras2019explaining}. State-of-the-art approaches involve using techniques such as model visualization, layer-wise relevance propagation, and attention mechanisms \cite{guo2019exploring,ding2020interpretable,lim2021temporal}. While such techniques provide valuable insights into the working of the model, extraction of physical knowledge about the system being modeled is limited. One step in this direction is to use ML models for extracting the timescales of the physical system, which are fundamental to understanding complex dynamical systems.

In this paper, we report on the use of time-series ML to close the URANS equations for one of the simplest configurations of SST, namely, homogeneous and decaying turbulence.  The SST configuration and URANS equations are presented in \S\ref{sec:theory}. In \S\ref{sec:approach}, we review time-series ML approaches, the training data sets, and techniques for interpreting the data requirements of the ML models. The results from our URANS models are discussed in \S\ref{sec:results}. Some conclusions are drawn in \S\ref{sec:conclusion}.


\section{Theoretical Background for URANS of SST}
\label{sec:theory}

For this work, we consider temporally evolving SST that is axisymmetric in the sense that the statistics are independent of horizontal direction.  The flow is stably stratified by a uniform density gradient so that it is homogeneous in the vertical direction.  While the configuration is formally homogeneous in all directions, we show in \S\ref{subsec:training_data} with contour plots from direct numerical simulations (DNS) that the flow is intermittent at length scales that are a significant fraction of the domain size. We take advantage of the homogeneity to reduce the RANS grid down to one point in each direction in \S\ref{subsec:training_data}.  Here, let us write the governing equations and their Reynolds-averaged versions for the general case.

The modeled flow evolves with density stratification and is governed by the dimensionless Navier--Stokes equations satisfying the Boussinesq assumption, given by
\vspace{-\baselineskip}

\begin{align}
    \nabla \cdot \boldsymbol{u} &= 0, \label{eq:NS-continuity}\\
    \frac{\partial \boldsymbol{u}}{\partial t} + \boldsymbol{u} \cdot \nabla \boldsymbol{u} &= -\left( \frac{2\pi}{Fr} \right)^2 \rho \boldsymbol{e}_z -\nabla p + \frac{1}{Re}\nabla^2\boldsymbol{u}, \label{eq:NS-momentum}\\
    \frac{\partial \rho}{\partial t} + \boldsymbol{u} \cdot \nabla \rho - w &= \frac{1}{Re Pr} \nabla^2 \rho. \label{eq:NS-density}
\end{align}
Here, $\boldsymbol{u} = (u,v,w)$ represents the velocity vector in the Cartesian coordinate system $\boldsymbol{x} = (x,y,z)$ evolving in time $t$, and $\rho$ and $p$ are the deviations of density and pressure from their ambient conditions, respectively. The velocity and length scales for non-dimensionalzing the variables are the r.m.s. velocity $\hat{U}$ and integral length scale $\hat{L}$, respectively. Note that the dimensional quantities are represented with $\hat{\cdot}$. Time $t$ is non-dimensionalized as $t = (\hat{t} - \hat{t}_0)/\hat{t}_{LE}$, where $\hat{t}_0$ is the initial time at which the turbulence decay process starts and $\hat{t}_{LE}$ is the large-eddy advective time scale. The non-dimensional quantities, Froude number $Fr$, Reynolds number $Re$, and Prandtl number $Pr$, are defined respectively as
\begin{equation}
    Fr = \frac{2\pi \hat{U}}{\hat{N}\hat{L}}, Re = \frac{\hat{U}\hat{L}}{\hat{\nu}}, ~\text{and}~ Pr=\frac{\hat{\nu}}{\hat{\alpha}}.
\end{equation}
Here, $\hat{N} = \left[-(\hat{g}/\hat{\rho}_0) (\text{d}\overline{\hat{\rho}}/\text{d}\hat{z})\right]^{1/2}$ is the Brunt--V\"{a}is\"{a}l\"{a} frequency or the buoyancy frequency, $\hat{g}$ is the acceleration due to gravity in the $-z$ direction, $\hat{\rho}_0$ is the density at a reference height of $\hat{z}_0$, $\hat{\nu}$ is the kinematic viscosity, and $\hat{\alpha}$ is the thermal diffusivity. The density gradient $\text{d}\overline{\hat{\rho}}/\text{d}\hat{z}$ is a uniform, time-invariant, stable ambient density gradient applied to impose stratification.

\begin{figure}
\begin{center}
  \includegraphics[width=1\textwidth]{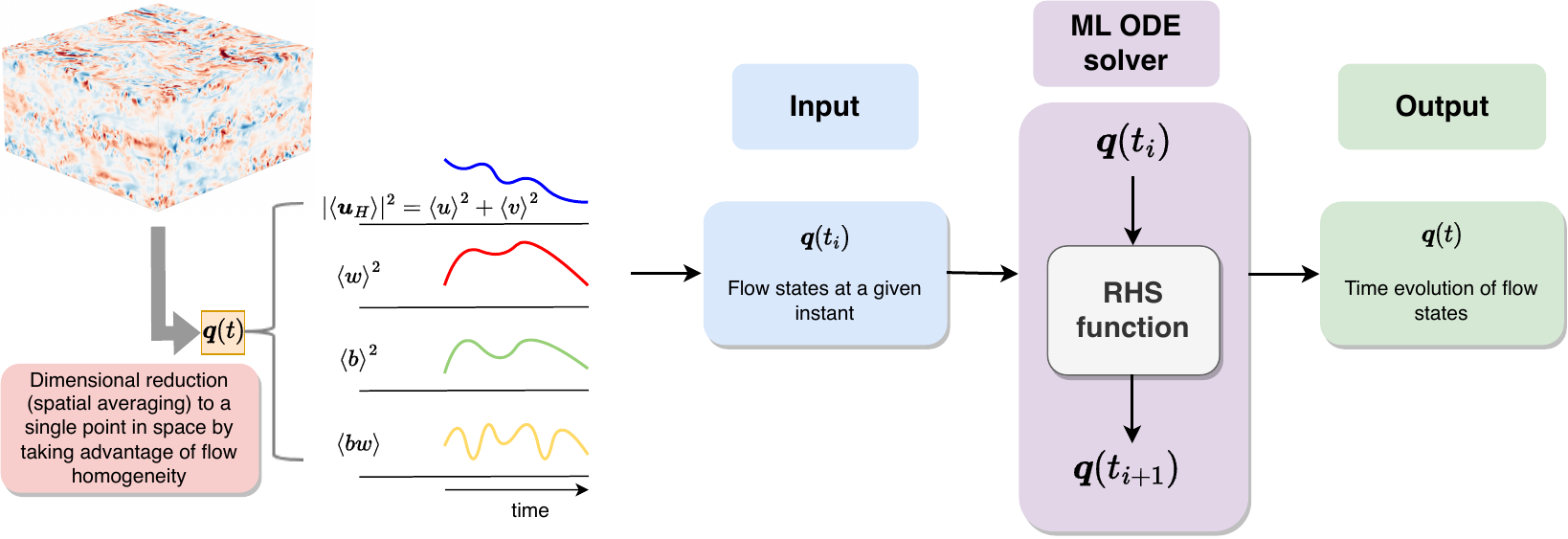}
\end{center}
 \caption{Framework of the current modeling approach to predict the time evolution of the reduced-dimensional flow state variables.}
\label{fig:framework}
\end{figure}

For the homogeneous case, we can approximate the Reynolds average with a spatial average over the entire domain, denoted by $\left< \cdot \right>$. Then the time evolution of the average flow can be described in terms of energy equations by decomposing kinetic energy into its horizontal and vertical contributions, $E_H$ and $E_V$, which are coupled via pressure. The potential energy $E_P$ is a scalar multiple of the variance of the buoyancy. An additional equation is required for the buoyancy flux, which couples $E_V$ and $E_P$. The resulting set of evolution equations is
\vspace{-\baselineskip}

\begin{align}
    \frac{\text{d}}{\text{d}t} \left<E_H\right> &= \underbrace{\left< p \nabla_H\cdot\boldsymbol{u}_H \right>}_\text{needs model} - \underbrace{\hat{\nu} \left< \left( \frac{\partial \boldsymbol{u}_{H}}{\partial x_j} \right)^2 \right>}_\text{needs model}\label{eq:RANS-Eh}\\
    \frac{\text{d}}{\text{d}t} \left<E_V\right> &= \underbrace{\left< p \frac{\partial w}{\partial z} \right>}_\text{needs model} - \left< bw \right> - \underbrace{\hat{\nu} \left< \left( \frac{\partial w}{\partial x_j} \right)^2 \right>}_\text{needs model}\label{eq:RANS-Ev}\\
    \frac{\text{d}}{\text{d}t} \left<E_P\right> &= \left< bw \right> - \underbrace{\frac{\hat{\alpha}}{\hat{N}^2}\left< \left( \frac{\partial b}{\partial x_j} \right)^2 \right>}_\text{needs model}\label{eq:RANS-Ep}\\
    \frac{\text{d}}{\text{d}t} \left<bw\right> &= \underbrace{\left< p \frac{\partial b}{\partial z} \right>}_\text{needs model} + \left< w^2 \right>\hat{N}^2 - \left< b^2 \right> - \underbrace{\left( \hat{\nu} + \hat{\alpha} \right) \left< \left( \frac{\partial b}{\partial x_j} \right) \left( \frac{\partial w}{\partial x_j} \right) \right>}_\text{needs model}\label{eq:RANS-bw}
\end{align}
where $E_H = (1/2)|\boldsymbol{u}|^2_H$, $E_V = (1/2)w^2$, and $E_P = b^2 / (2\hat{N}^2)$. Here, $b=\frac{\hat{g}}{\hat{\rho}_0}\hat{\rho}$ is the buoyancy, and $|\boldsymbol{u}_H| = \left( u^2 + v^2 \right)^{1/2}$ is the horizontal velocity. Several terms on the right-hand-sides (RHSs) of these equations are marked to indicate that they must be modeled in order to close the system of ordinary differential equations. Rather than model each term, or even the RHS of each equation, our objective is to model the entire set of RHSs using ML techniques, as broadly depicted in \fig\ref{fig:framework}.

\section{Approach}
\label{sec:approach}

\subsection{Procedure}\label{subsec:procedure}

As mentioned in \sectn\ref{sec:theory}, the objective is to model the temporal evolution of the energies (horizontal and vertical kinetic energies, and potential energy) and buoyancy flux for a single point in space. Approximating the derivatives of the energy variables are of interest for modeling the RHS of \eqns\ref{eq:RANS-Eh}-\ref{eq:RANS-bw}, needed to close the system. The terms to be modeled can be represented as
\begin{equation}
    \boldsymbol{f} = \left[\frac{\text{d}}{\text{d}t} \left<E_H\right> \quad \frac{\text{d}}{\text{d}t} \left<E_V\right> \quad \frac{\text{d}}{\text{d}t} \left<E_P\right> \quad \frac{\text{d}}{\text{d}t} \left<bw\right>\right]^T.
    \label{eq:nn_output}
\end{equation}
The terms are non-dimensionalized by the total energy, $E_T = \left<E_H\right>+\left<E_V\right>+\left<E_P\right>$, and buoyancy frequency, $\hat{N}$, at the initial condition, giving $\left<E_H\right> = (1/2)|\left<\boldsymbol{u}_H\right>|^2/E_{T_0}$, $\left<E_V\right> = (1/2)\left<w\right>^2/E_{T_0}$, $\left<E_P\right> = \left<b\right>^2 / (2\hat{N}^2E_{T_0})$, and $\left<bw\right>/(2\hat{N}E_{T_0})$ as the terms to be modeled. Here, we can represent the independent variables as $\boldsymbol{q} = \left[ \left|\left<\boldsymbol{u}_H\right>\right|^2 \quad \left<w\right>^2 \quad \left<b\right>^2 \quad \left<bw\right> \right]^T$. Thus, the problem can be generically represented in discrete time as
\begin{equation}
    \frac{\text{d}\boldsymbol{q}(t_i)}{\text{d}t} = \boldsymbol{f}\left(\boldsymbol{q}(t_i), t_i\right)
\label{eq:generic_ode}
\end{equation}
where $t_i$ represents a discrete time instant. We use supervised deep learning techniques \cite{lecun2015deep} to model the relationship between $\boldsymbol{q}$ and $\boldsymbol{f}$, and thus, model the time evolution of $\boldsymbol{q}$, as shown in \fig\ref{fig:framework}.

Neural networks have the capacity to approximate any Borel measurable function if it has the proper architecture \cite{hornik1989}. We are interested in neural networks capable of modeling the RHS of the system of ODEs in \eqn\ref{eq:generic_ode}, representing the reduced form of \eqns\ref{eq:RANS-Eh}-\ref{eq:RANS-bw}, so that any numerical ODE solver can solve the time evolution of the states, $\boldsymbol{q}(t)$. 
Moreover, as the problem is fundamentally dependent on time and the time history of the states, modeling the \textit{memory effect} present in the time series using time-series ML models is attractive. Thus, our objective revolves around using sequences of time-series data to predict the time evolution of $\boldsymbol{q}$. Here, sequence refers to the ordered snapshots of state time history (or \textit{memory}) used to model the dynamics of the system.

\begin{figure}
\begin{center}
  \includegraphics[width=0.4\textwidth]{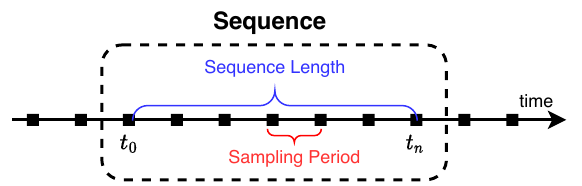}
\end{center}
\caption{Illustration of the timescales defined by a sequence in a time-series data.}
\label{fig:ml_timescales}
\end{figure}

Generally, sequences of time-series data have two parameters analogous to timescales in physical time-evolving systems, as illustrated in \fig\ref{fig:ml_timescales}. One timescale is the sequence length, which is the time span of the data in a single sequence. The second timescale is the sampling period of the data within one sequence. We will refer to latter timescale by its inverse, the sampling frequency. With respect to physical timescales, the sequence length can be considered as the largest timescale of the system being modeled whereas the sampling period can be treated as the shortest timescale of the system. We will elaborate more on what these timescale mean for SST in \S\ref{sec:results}.

We use these underlying timescales of the temporal data as hyperparameters of time-series ML models. Furthermore, we determine the minimum information, based on these timescales, required by the  ML models to effectively model the system dynamics. These techniques are used specifically for interpreting the data requirements for the time-series ML models.

We choose to model the evolution of the flow states using two routes: (1) model the time derivatives (or the entire RHS of \eqns\ref{eq:RANS-Eh}-\ref{eq:RANS-bw}) using an ML model and pass this information into an ODE solver to obtain the time evolution of $\boldsymbol{q}$ and (2) directly model the time evolution of $\boldsymbol{q}$ using an ML framework.
We achieve these two routes using the Long Short-Term Memory (LSTM) \cite{hochreiter1997long} and Neural Ordinary Differential Equation (NODE) \cite{chen2018neural} models, respectively.
In the following sections, we describe the architecture of the LSTM and NODE models, and the dataset used to train these models.

\subsection{Neural network setup}\label{subsec:nn_setup}

Without a loss of generality, the system of differential equations governing the SST, \eqn\ref{eq:generic_ode}, are modeled as a neural network (NN), $\mathcal{N}$, given by
\begin{equation}
    \frac{\text{d}\boldsymbol{q}(t_i)}{\text{d}t}=\underbrace{\tilde{\boldsymbol{f}}(\boldsymbol{q}(t_i),t_i,\theta)}_\text{$\mathcal{N}$}
    \label{eq:generic_ode_ML}
\end{equation}
where the NN is parameterized by a tensor, $\theta$, containing learnable weights and biases defining the map between the input and output layers of the network. The symbol $\tilde{\cdot}$ represents the quantities obtained from the learned ML model.

\subsubsection{Long Short-Term Memory (LSTM) model}

\begin{figure}
\begin{center}
  \includegraphics[width=0.6\textwidth]{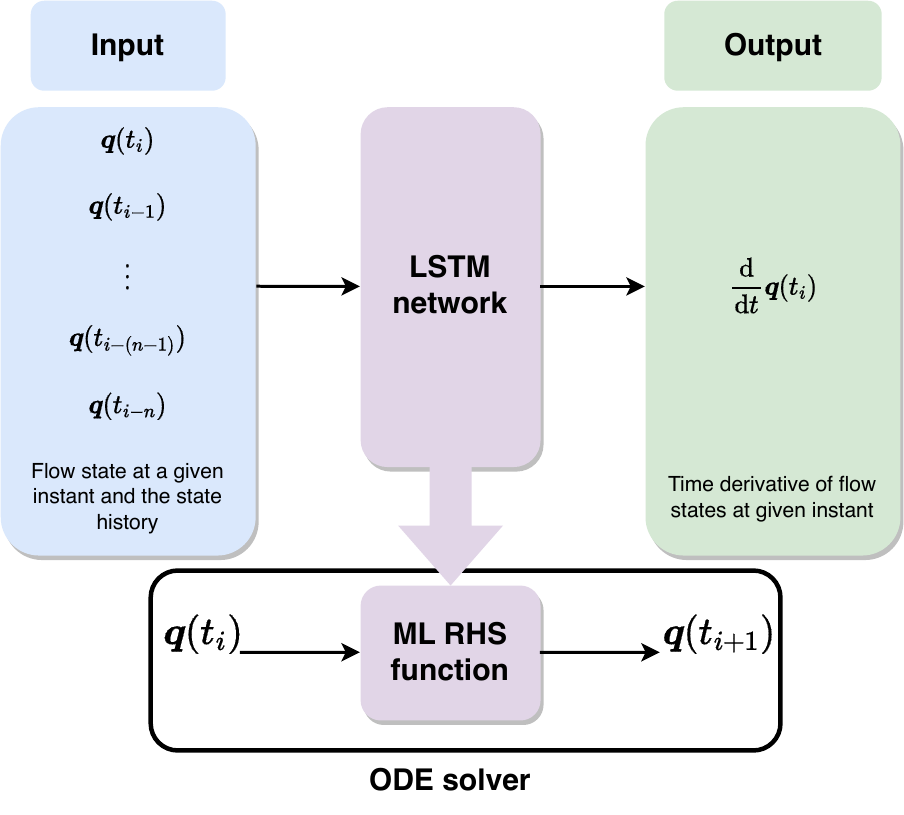}
\end{center}
\caption{Sample portrayal of the LSTM modeling framework used in the current study for predicting the time evolution of a system of ODEs.}
\label{fig:framework-LSTM}
\end{figure}

In the first route for modeling the system of ODEs, given the components of $\boldsymbol{q}$ at an instant in time $t_i$, we use NNs of the form $\mathcal{N}(\boldsymbol{q}, \theta)$ to model the RHS of \eqn\ref{eq:generic_ode_ML}, with $\theta$ representing the parameters of the NN (``weights''). Specifically, we use the Long Short-Term Memory (LSTM) \cite{hochreiter1997long} network as the NN model. An LSTM architecture is a sub-class of the recurrent neural network (RNN) architecture, which uses internal states of nodes (memory) to process arbitrary sequences of inputs and thus, allows for time-series to be modeled \cite{connor1994recurrent,che2018recurrent,hewamalage2021recurrent}. LSTM is commonly used for time-series prediction because of its improved stability and predictive capabilities over generic RNNs \cite{hochreiter1997long}. A sample depiction of the model is shown in \fig\ref{fig:framework-LSTM}.

The input to the LSTM model is the time history $\boldsymbol{q}(t_{i-n})$, $\boldsymbol{q}(t_{i-(n-1)})$, \ldots, $\boldsymbol{q}(t_i)$ where $n$ is the sequence length. The output of LSTM is $\tilde{\boldsymbol{f}}\left(\boldsymbol{q}(t_i), t_i,\theta\right)$, which is fed into an ODE solver (like a Runge--Kutta scheme) to obtain the next value in the series, $\boldsymbol{q}(t_{i+1})$. The process is repeated to predict the entire time evolution of $\boldsymbol{q}$. Thus, the entire workflow can be viewed as a LSTM$+$ODE numerical simulation, where the LSTM serves as the RHS of the ODE solver. The LSTM model is trained using the loss function comprised of the RHS of \eqn\ref{eq:generic_ode_ML}, given by
\begin{equation}
    \mathcal{L}\left[\boldsymbol{f}\left(\boldsymbol{q}(t_i),t_i\right), \tilde{\boldsymbol{f}}\left(\boldsymbol{q}(t_i),t_i,\theta\right)\right].
\end{equation}

Hyperparameter selection for the LSTM model is done through experimentation (broad grid search). We found that for modeling the SST problem, the best predictions are obtained with four LSTM layers with a hidden size of 10. We also use a multi-layer perceptron (MLP) consisting of one hidden layer with 15 neurons before the output layer. A \textit{Leaky} Rectified Linear Unit (LeakyReLU) activation function with a slope of $0.1$ is applied for both the LSTM and MLP layers. We measure the mean squared error as the loss function, $\mathcal{L}$, for optimizing the network weights during the training process. We select a learning rate of $0.001$ and train the model for $4000$ epochs, after which the model training converges.

\subsubsection{Neural Ordinary Differential Equation (NODE) model}

In the second route for modeling the system of ODEs, given the components of $\boldsymbol{q}$ at an instant in time $t_i$, we use an ML framework devised by \citet{chen2018neural}, Neural Ordinary Differential Equations (NODE), to directly predict the time evolution of the system. The NODE framework models an arbitrary system of ODEs as an NN given discrete observations. \citet{legaard2023constructing} refer to this framework as a time-stepper model compared to direct solution models which map a given time, $t_i$, to its observation, $\boldsymbol{q}_i$, such as deep neural networks (DNN), recurrent neural networks (RNN), and residual neural networks (ResNet) \cite{he2016deep}. Motivated by the similarity between ResNets and Euler's method for numerical integration, wrapping a black-box ODE solver around an NN offers many modeling advantages. The main advantage being that, unlike ResNets or traditional sequential learning architectures, transformations between hidden states evolve continuously. Many systems described by differential equations evolve as such, so the combination of well known mathematical integration tools and neural networks can be a powerful paradigm in scientific applications. In recent years, NODE has been used to model complex systems in fluid dynamics, such as to model and predict thermoacoustic instability \cite{dhadphale2022neural} and learn subgrid-scale models \cite{kang2023learning}. Related work also includes the implementation of ODE-based ML solutions for modeling dynamical systems \cite{linot2023stabilized}, reduced-order modeling \cite{linot2022data}, and turbulent flow control \cite{zeng2022data,linot2023turbulence}.

\begin{figure}
\begin{center}
  \includegraphics[width=0.6\textwidth]{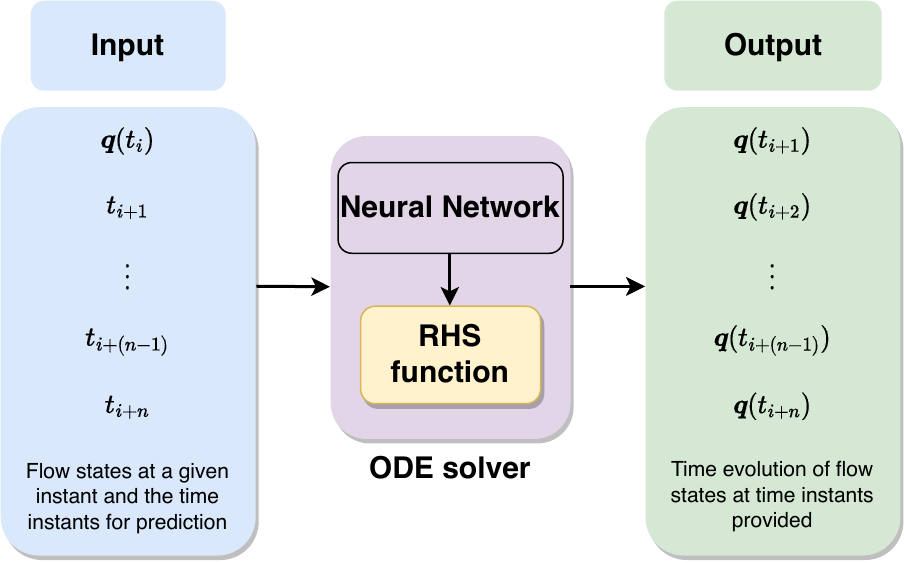}
\end{center}
\caption{Sample portrayal of the NODE modeling framework used in the current study for predicting the time evolution of a system of ODEs.}
\label{fig:framework-NODE}
\end{figure}

As depicted in \fig\ref{fig:framework-NODE}, NODE takes as input the current state of the system $\boldsymbol{q}(t_i)$ and the desired time instants $t_{i+1}$, $t_{i+2}$, \ldots, $t_{i+n}$ at which to evaluate the state. The output is the state variables $\boldsymbol{q}(t_{i+1})$, $\boldsymbol{q}(t_{i+2})$, \ldots, $\boldsymbol{q}(t_{i+n})$ at those discrete time instants. What makes a time-stepper model unique compared to direct solutions, is that the loss is computed after integrating the learned approximation of the differential equation, \eqn\ref{eq:generic_ode_ML}. As a result, the input to the scalar loss function $\mathcal{L}$ is the output of the ODE solver, given by
\begin{equation}
    \mathcal{L}\left[\boldsymbol{q}(t_{i+n}), \tilde{\boldsymbol{q}}(t_{i+n})\right] = \mathcal{L}\left[\boldsymbol{q}(t_{i+n}), \boldsymbol{q}(t_i)+  \int_{t_i}^{t_{i+n}} \tilde{\boldsymbol{f}}(\boldsymbol{q}(t),t,\theta)\text{d}t\right] = \mathcal{L}\left[\boldsymbol{q}(t_{i+n}), ODESolve(\boldsymbol{q}(t_i), \tilde{\boldsymbol{f}},t_i,t_{i+n},\theta)\right]
\end{equation}
where $t_i$ and $t_{i+n}$ represents the initial and final time instants for a given sample of the time-series data. Gradients can be computed using the adjoint sensitivity method, solving an augmented ODE backwards in time. The loss function can then be minimized by comparing the integrated prediction, $\tilde{\boldsymbol{q}}(t)$, and observed solution, $\boldsymbol{q}(t)$, in relation to the tensor, $\theta$, which parameterizes the differential equation. Using this framework, recovering both the approximation of the state variables and their derivatives are convenient. Derivatives are recovered by inputting the predictions back into the neural network defining the temporal derivative of the system, $\tilde{\boldsymbol{f}}(\boldsymbol{q}(t),t,\theta)$ (\eqn\ref{eq:generic_ode_ML}).

In literature, NODE has been shown to be less sensitive to hyperparameter selection compared to traditional NN models, like MLPs and ResNets \cite{rahman2022}. This pertains to the hyperparameters of the model architecture and not the two \textit{data hyperparameters} of interest in the current study -- sequence length and sampling frequency. For the \textit{model hyperparameters} of the NODE architecture, we chose an absolute tolerance of $10^{-3}$ and a relative tolerance of $10^{-4}$ for the integrated ODE solver \cite{chen2018neural}. Furthermore, the MLP within the NODE model has 10 layers with $40$ neurons per layer. We use a Sigmoid Linear Unit (SiLU) activation function for these layers. Mean squared error is used as the loss function, $\mathcal{L}$, for optimizing the network parameters, $\theta$, during the training process. For this configuration, we used a higher learning rate of $0.05$ compared to LSTM and lower number of epochs of $1000$ to train the NODE model. This is because the NODE model converged faster with similar performance.

\subsubsection{Comparing LSTM and NODE models}

\begin{figure}
\begin{center}
  \includegraphics[width=0.7\textwidth]{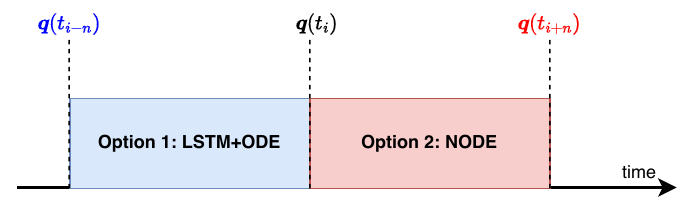}
\end{center}
 \caption{Comparing the time-dependence for the LSTM and NODE models.
\label{fig:time-LSTM-NODE}}
\end{figure}

As described above, the time-series prediction that we formulate the LSTM and NODE models to perform are fundamentally different. In general, the inputs and outputs of the LSTM and NODE models can be summarized as shown in \tbl\ref{table:model-io} and their time-dependence can be summarized as shown in \fig\ref{fig:time-LSTM-NODE}. While the LSTM model inherently relies on the time history of the data, the NODE model utilizes the future time evolution of the data to model the behaviour of the system. Nonetheless, there are two timescales for the input data to both the models - the sequence length and sampling frequency. Using these two timescales as the main hyperparameter of the two models, we investigate the minimum information (with respect to these two timescales) that is required to effectively model a system of ODEs - assessed in terms of the accuracy and numerical stability of the resulting ODE solver.

\begin{table}
\centering
\caption{Inputs and outputs of the LSTM and NODE models.}
\begin{tabular}{ccc}
 & LSTM & NODE  \\
 \hline\\
\textbf{Inputs} & $\boldsymbol{q}(t_{i-n}), \boldsymbol{q}(t_{i-(n-1)}), \dots, \boldsymbol{q}(t_{i})$ & $\boldsymbol{q}(t_i)$, $t_{i}$, $t_{i+1}$, \dots , $t_{i+n}$  \\\\
\textbf{Outputs} & $\boldsymbol{f}(t_i)$ & $\boldsymbol{q}(t_{i+1})$, $\boldsymbol{q}(t_{i+2})$, \dots, $\boldsymbol{q}(t_{i+n})$
\end{tabular}
\vspace{0.02\textwidth}
\label{table:model-io}
\end{table}

\subsection{Training the models}\label{subsec:training_data}

\begin{figure}
\begin{center}
  \includegraphics[width=1\textwidth]{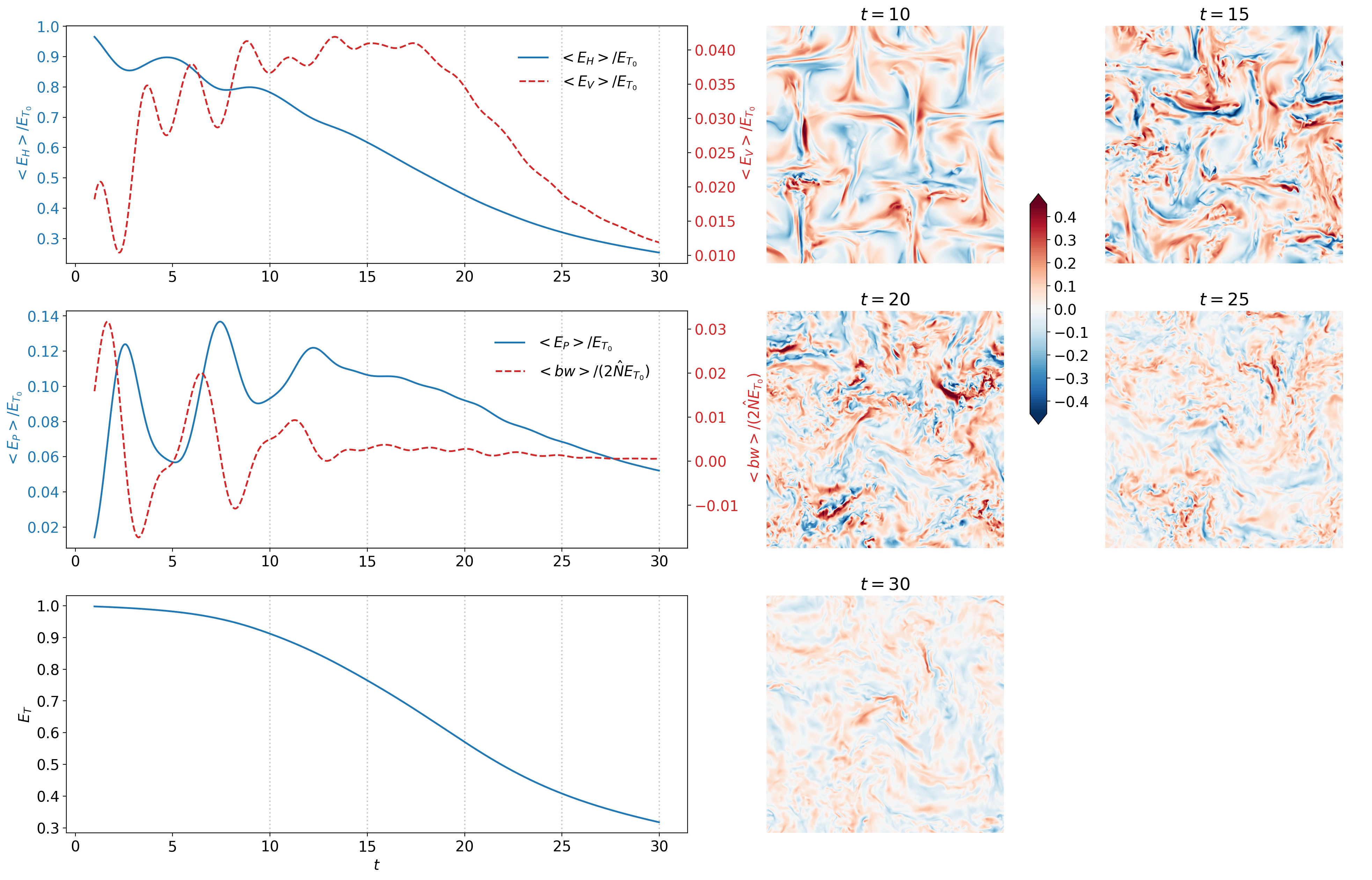}
\end{center}
 \caption{Time series data of the SST flow being modeled, with candidate horizontal slices of vertical velocity, $w$, at various time instants.
\label{fig:F4R32-sample}}
\end{figure}

The time-series models are trained using data from a decaying homogeneous SST flow simulation described in \citet{riley2003dynamics}. A sample portrayal of the time evolution of the SST flow is shown in \fig\ref{fig:F4R32-sample}. The flow is initialized using Taylor--Green vortices with low-level noise ($10\%$ of initial energy) to break the symmetries of the vortices and trigger instabilities for the flow to evolve into turbulence. The ambient stratification is constant and the initial density perturbation field is zero. This configuration is a significant test for URANS because it starts out laminar, then transitions to turbulence, and finally becomes increasingly more strongly stratified as it decays such that the inertial force decreases relative to the buoyancy force.   We consider the case with initial Reynolds and Froude numbers $Re=3200$ and $Fr=4$, which is discussed in detail in \citep{hebert06a,hebert06b}.

As mentioned in \sectn\ref{sec:theory}, we take advantage of the homogeneity of the flow and spatially average the spatio-temporal data to a single grid point and obtain time series of the state variables. The training data consist of sequence of these time-series data, defined by two time scales: (1) the length of the sequence, sequence length, and (2) the time interval between samples within the sequence, the sampling frequency. We use these two variables as hyperparameters to interpret the data requirements of the models, where each model is trained for a single type of sequence defined by the hyperparameters. Note that these \textit{data hyperparameters} are different from the \textit{model hyperparameters}, which are chosen as mentioned in \sectn\ref{subsec:nn_setup} and kept constant throughout the analysis. For each hyperparameter set, the time-series is interpolated to have the same time-step as that of the sampling frequency. Thus, the number of samples used for training the models is determined by the sequence length as well as the sampling frequency. 

For a given set of samples defined by the \textit{data hyperparameters}, we split the time-series to training and testing regimes based on the temporal evolution of the system. We chose to train the models untill $t=20$, as the data captures both the transient as well as the beginning of the turbulent decay regimes. The testing is done both for the unseen data as well as from the beginning (in the training regime) to test both the accuracy and numerical stability of the model. The full training data for a given set of hyperparameters is split with a ratio of $90:10$ for training and validation at each training epoch. As the terms modeled are already non-dimensionalized, we did not have to perform additional scaling of the data, which is typical for NN training due to the limited range in which activation functions can operate.


\section{Results and Discussion}
\label{sec:results}

Here, we evaluate the ability of time-series ML to learn the dynamics of the reduced-dimensional SST by modeling the system of equations governing the force balance of SST URANS equations. We demonstrate that data requirements of these models are interpretable by extracting the physical timescales corresponding to the information learned by the model. We study the relationship between the physical timescales and the ML hyperparameters analogous to timescales of the time-series data -- the \textit{data hyperparameters} sequence length and sampling frequency. 

To demonstrate the interpretability framework, we first model the time evolution of a damped pendulum \cite{nelson1986} as a sample model problem. The damped pendulum is a very simple and well-studied nonlinear dynamical system. It has two timescales that can be derived analytically, which makes it a suitable choice for this study. Although the damped pendulum problem is not as complex as the nonlinear SST that motivates this study, we use the damped pendulum problem as a canonical model problem to demonstrate our framework of extracting physical timescales from time-series ML models \cite{liousas2024predictinging}. Herein, we first show the results for the \textit{data hyperparameters} (sequence length and sampling frequency) for which we obtained the best results and then we discuss the effects of these hyperparameters on the model performance and how they can be used to interpret the data requirements of the models.

\subsection{Sample test problem: Simple pendulum}

Machine learning models for both the LSTM and NODE are tested and validated with a simpler, well studied, synthetic dataset describing the motion of a damped pendulum. A damped pendulum exhibits some of the fundamental behaviours observed in the results for homogeneous SST, making it a relevant dataset for evaluating model performance. The governing equations of the pendulum are typically written in terms of its angular position $\theta$ and angular velocity $\omega$. However, many dynamical systems in engineering and nature can be described by energy equations. In particular, the motivation for this study, URANS modeling of SST, is formulated in terms of energies as described in \eqns\ref{eq:RANS-Eh}-\ref{eq:RANS-bw}. The system of ODEs governing the time evolution of the pendulum in terms of its kinetic energy ($E_K=\frac{1}{2}m\ell^2\omega^2$), potential energy ($E_P=mg\ell(1-\cos\theta)$), and the flux ($F=mg\ell\omega\sin\theta$) between them is given by
\vspace{-\baselineskip}

\begin{align}
    \frac{\text{d}E_K}{\text{d}t} &= -\frac{2 b}{m} E_K - F \label{eq:dK} \\
\frac{\text{d}E_P}{\text{d}t} &= F \label{eq:dP} \\
\frac{\text{d}F}{\text{d}t} &= 2\frac{g}{\ell}(E_K-E_P) + \frac{E_P}{m \ell^2}(2E_K - E_P) -\frac{b}{m}F \label{eq:dF},
\end{align}
where the parameters of the pendulum are the damping coefficient $b$, mass $m$, acceleration due to gravity $g$, and length $\ell$. Thus, the state of the system is defined as $\boldsymbol{q} = \left[E_K \quad E_P\quad F\right]^T$. Solving the system of ODEs requires the initial conditions $E_{K_0}$, $E_{P_0}$, and $F_0$. The similarities of the states of the pendulum and the SST problem are notable.

The pendulum has two fundamental timescales that govern its motion. They are the oscillating timescale $\tau_o$ and the damping timescale $\tau_d$. These can be derived analytically using the small-angle approximation \cite{nelson1986} and are given by
\vspace{-\baselineskip}

\begin{align}
    \tau_o &= \sqrt{\frac{\ell}{g}} \label{eq:tau_o} \\
    \tau_d &= \frac{m}{b} \label{eq:tau_d}.
\end{align}

We can use these timescales to determine the theoretical minimum information that a data-driven model would need to effectively learn the dynamics of a pendulum. An accurate model must capture the information corresponding to both the damping as well as oscillating timescales.

\subsubsection{LSTM and NODE results}\label{subsubsec:pendulum-results}

We select a damped pendulum with damping coefficient $b=0.5$, mass $m=1$, acceleration due to gravity $g=25$, length $\ell=l$, and initial conditions $E_{K_0}=0$, $E_{P_0}=1$, and $F_0=0$. Note that for these parameters, $\tau_d >> \tau_o$. Thus, the pendulum is underdamped and oscillates considerably before damping, similar to SST. The data for training and testing the models were generated numerically using these settings of the pendulum. Training and testing of the models were performed at different instants in time.

\begin{figure}
\begin{center}
  \includegraphics[width=1\textwidth]{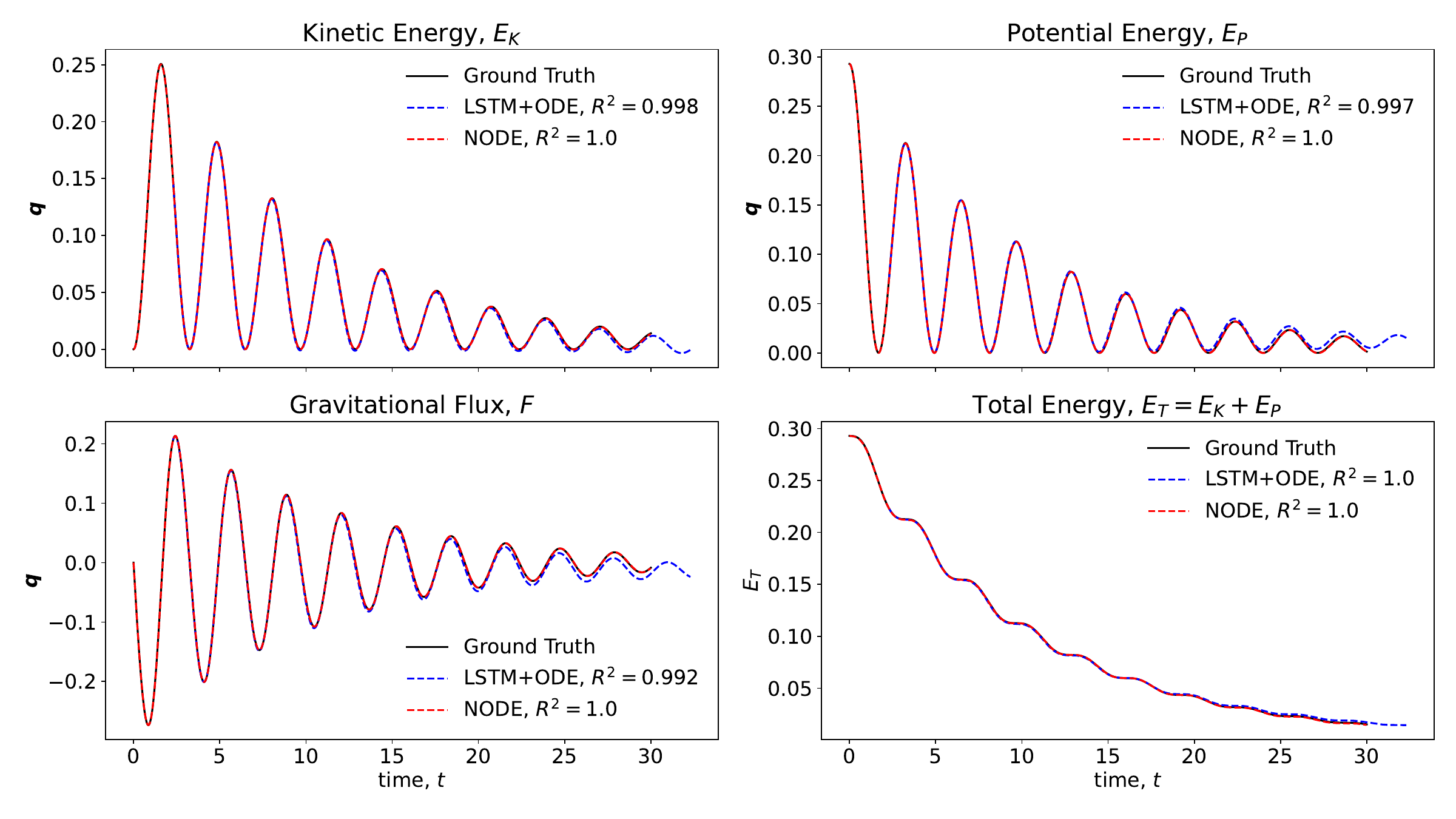}
\end{center}
 \caption{Comparison of the true (analytical) time evolution of the state variables of the damped pendulum with the predictions of the LSTM and NODE models. The error between the true values and ML models is computed using the coefficient of determination, denoted by $R^2$.
\label{fig:results-pendulum}}
\end{figure}

We show the results for these best LSTM and NODE models in \fig\ref{fig:results-pendulum}. Both models perform accurately and are numerically stable. We observe that the NODE model slightly outperforms the LSTM model. It is interesting to note that the total energy of the pendulum $E_T$, \fig\ref{fig:results-pendulum} (bottom-right), is accurately replicated by both the LSTM and NODE models. $E_T$ was not explicitly modeled by the two time-series ML models, nor were any constraints imposed on the model training to follow the decay of $E_T$. The values for $E_T$ are computed post-simulation, by summing the kinetic and potential energies from the ML models.

\subsubsection{Interpreting the ML model data requirements: minimum information required by model}\label{subsubsec:pendulum-parametric}

The interpretability of the data requirements of the time-series ML models is studied by attempting to extract physically realizable timescales from the ML models. For this, the \textit{data hyperparameters}, sequence length and sampling frequency of the input data, are varied to see the effects on the model performance. The objective is to identify the minimum values of the hyperparamters which results in good model performance, thus finding the minimum information that is required by the ML models to effectively learn the system dynamics. 

\begin{figure}
\begin{center}
  \includegraphics[width=1\textwidth]{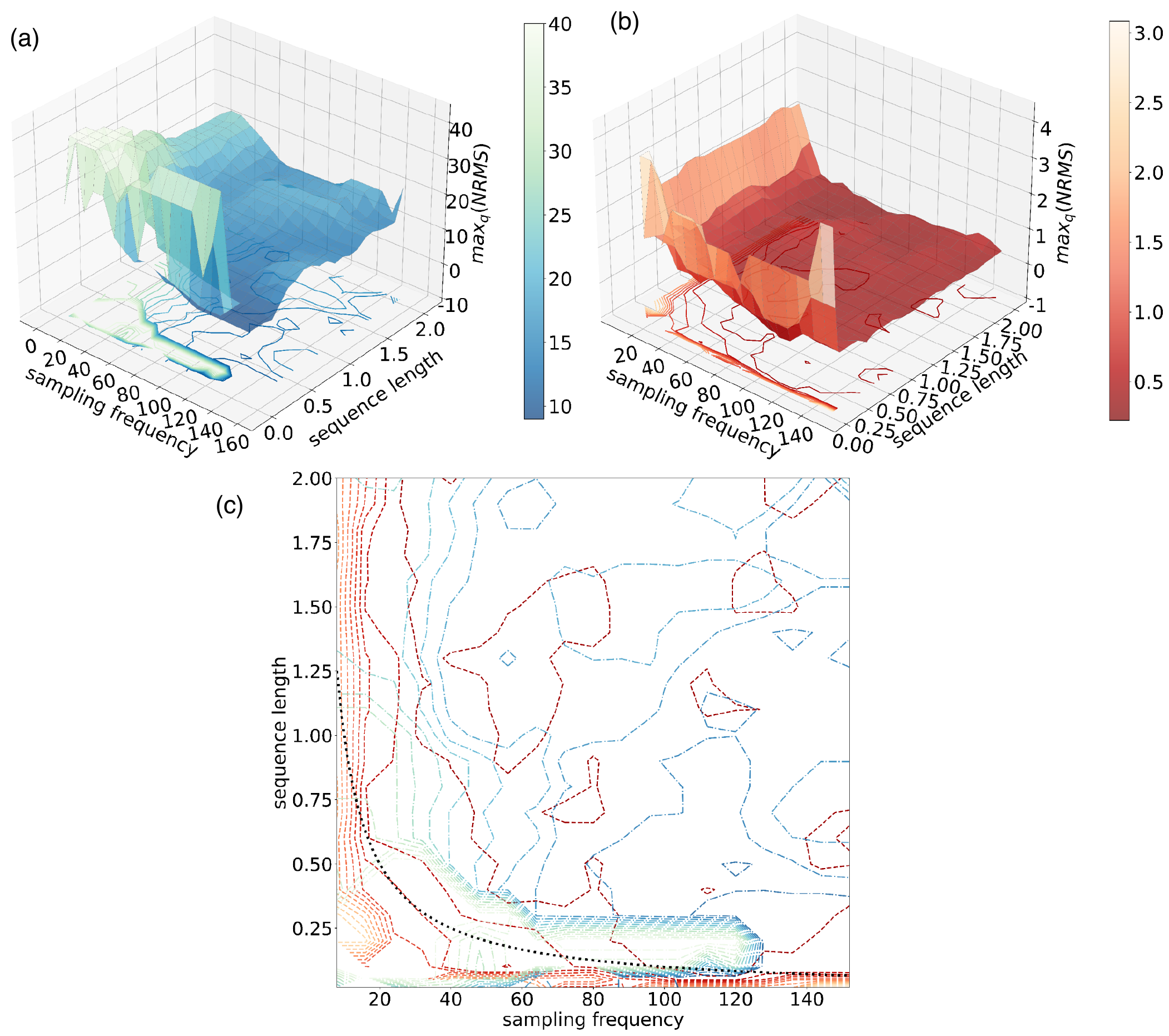}
\end{center}
 \caption{Effect of sequence length and sampling frequency of input data on (a) LSTM model and (b) NODE model. (c) Comparing the time-scales identified by LSTM (blue/green $-.$ contour) and NODE models (red/orange $--$ contour) based on effect of sequence length and sampling frequency with respect to the analytical timescales (black $\cdots$ line denoting minimum information) .
\label{fig:results-pendulum-parametric}}
\end{figure}

The results of the parametric study of the two ML timescales are shown in \fig\ref{fig:results-pendulum-parametric}. The \textit{data hyperparameters} are varied and the corresponding performance of the LSTM and NODE models are observed using 3-D contour plots (\fig\ref{fig:results-pendulum-parametric} (a-b)) and their corresponding 2-D projections (\fig\ref{fig:results-pendulum-parametric} (c-d)). The accuracy of the models (vertical axis in \fig\ref{fig:results-pendulum-parametric} (a-b) and the contour levels in \fig\ref{fig:results-pendulum-parametric} (c-d)) is measured by the maximum normalized root mean square error of all the states, $\max_{\boldsymbol{q}}(NRMS)$.

For the LSTM model, small values of sampling frequency and sequence length have extremely high errors as noted by the lower-left corner in \figs\ref{fig:results-pendulum-parametric} (a) and (c). This intuitively makes sense because when enough information is not provided to the model, the model is not able to learn the dynamics of the system. The 2-D projection of $\max_{\boldsymbol{q}}(NRMS)$ shown in \fig\ref{fig:results-pendulum-parametric} (c) reveals that this region lives below the minimum information line computed analytically (black dashed curve). The minimum information line corresponds to when the ratio of the sequence length and the sampling frequency is equal to the ratio of the damping timescale, $\tau_d$, and oscillating timescale, $\tau_o$, of the pendulum. 
Although there are local regions of high relative error, the general trend is that the LSTM model converges to accurate models beyond the minimum information required to represent the dynamics of the system. Thus, the inherent timescales of the minimum data input for accurate LSTM models corresponds to the physical timescale of the system.

The model error for the NODE models has a spike below the minimum information line similar to LSTM (lower-left corner of \figs\ref{fig:results-pendulum-parametric} (b) and (d)), whereas closely after the minimum information line, the NODE model performs accurately. Again, these results demonstrate that analyzing the inherent timescales of the input data to the time-series ML model can enable one to extract the timescales of the physical system. This ML-learned timescale correspond to the minimum information required by the time-series ML models to accurately predict the system dynamics.

The NODE model performs notably better than LSTM in the parametric study, as seen from \figs\ref{fig:results-pendulum-parametric} (a-d). Firstly, the highest error observed for the NODE model is an order of magnitude smaller than that of the LSTM model. Furthermore, NODE is much more numerically stable above the minimum information line since the surface plot is much flatter (and smoother) than that of LSTM, which has many local minima and maxima.

\subsection{SST initialized with Taylor--Green vortices}\label{subsec:SST-results}

We next demonstrate the capability of the time-series ML models to capture the nonlinear dynamics of the energy equations of the SST flow initialized with Taylor--Green vortices. We perform both \textit{a priori} (offline test) and \textit{a posteriori} (online test) testing of the LSTM and NODE models.

\subsubsection{\textit{a priori} testing}\label{subsubsec:SST-apriori}

\begin{figure}
\begin{center}
  \includegraphics[width=1\textwidth]{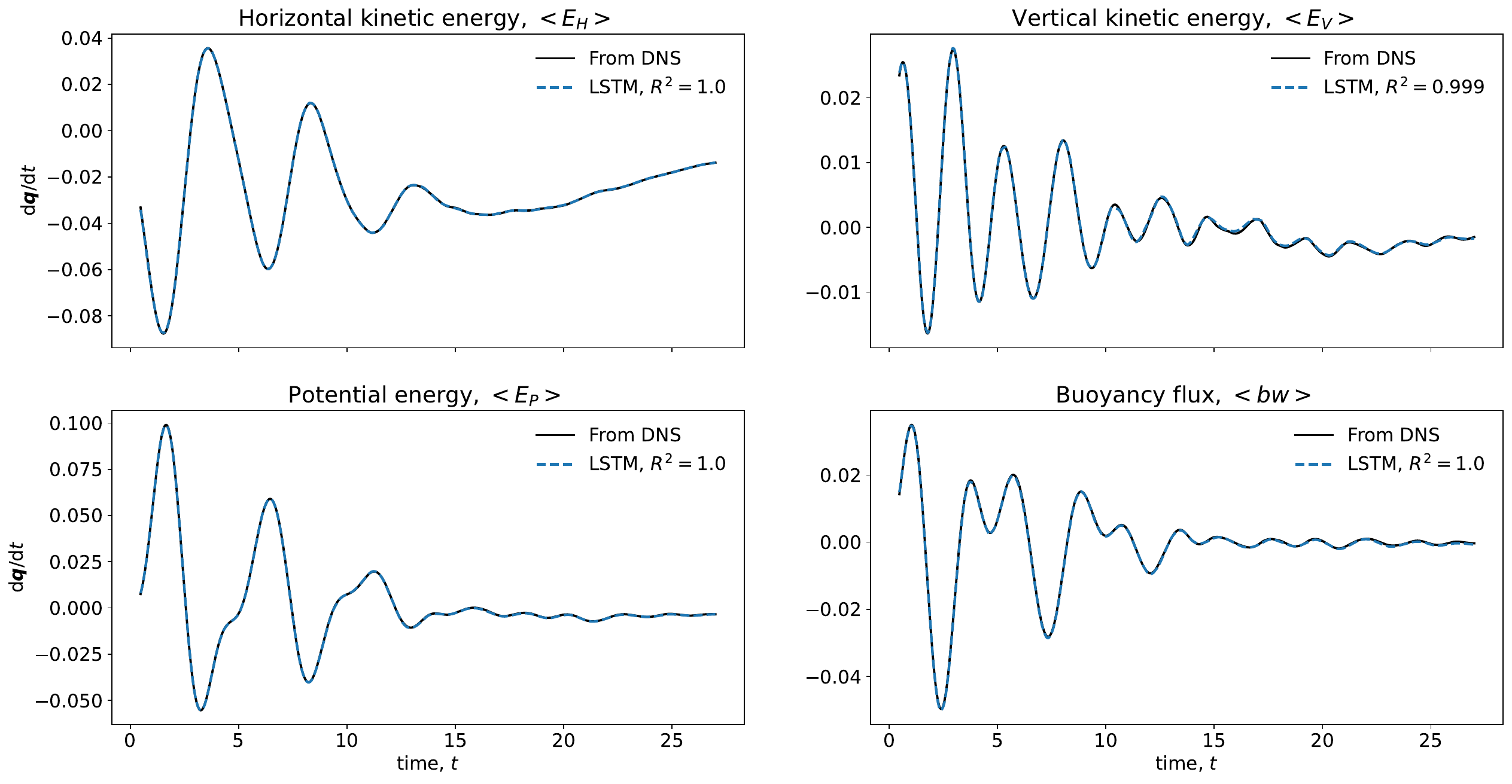}
\end{center}
 \caption{Results from \textit{a priori} tests using an LSTM model. The coefficient of determination, $R^2$, of the results compared to true outputs are also provided for each variable.
\label{fig:results-SST-apriori-LSTM}}
\end{figure}

The \textit{a priori} or offline test involves testing the accuracy of the models for a given set of inputs. For the LSTM model, we provide values of $\boldsymbol{q}$ at random time instants as inputs to the model and compare the outputs of the model with true or expected values -- the time derivative of $\boldsymbol{q}$ (i.e., ~$\boldsymbol{f}$) at the respective instants. The testing is performed at conditions within and outside the training data regime. The results of the \textit{a priori} test on the LSTM model are shown in \fig\ref{fig:results-SST-apriori-LSTM}, sorted in the order of the time evolution. We only show the results from the best model identified from the timescales analysis, which will be discussed more in \sectn\ref{subsubsec:SST-parametric}. The overall accuracy of the model for each variable is measured using the coefficient of determination, $R^2$, with respect to the true time derivatives of $\boldsymbol{q}$. The results demonstrates the high accuracy of the LSTM model to capture the complex nonlinear relationship between $\boldsymbol{q}$ and its time derivatives. For the NODE model, the outputs are the time evolution of $\boldsymbol{q}$ and thus corresponds to \textit{a posteriori} or online testing.

\subsubsection{\textit{a posteriori} testing}\label{subsubsec:SST-aposteriori}

The \textit{a posteriori} or online test involves testing the accuracy and stability of the models over the time evolution of the system. For the LSTM models, this involves coupling the output of the model with the simulations, which in this case is an ODE solver run using a fourth order explicit Runge--Kutta scheme. The outputs from the LSTM model, time derivatives of $\boldsymbol{q}$ at an instant (or the RHS of \eqn\ref{eq:generic_ode}, $\boldsymbol{f}$), serves as the RHS function for the ODE solver. We validate the numerical accuracy and stability of the ODE solver by comparing the time evolution of $\boldsymbol{q}$ obtained from DNS with that obtained from the ODE solver when using the true values of time derivative of $\boldsymbol{q}$ as the RHS of the solver.

\begin{figure}
\begin{center}
  \includegraphics[width=1\textwidth]{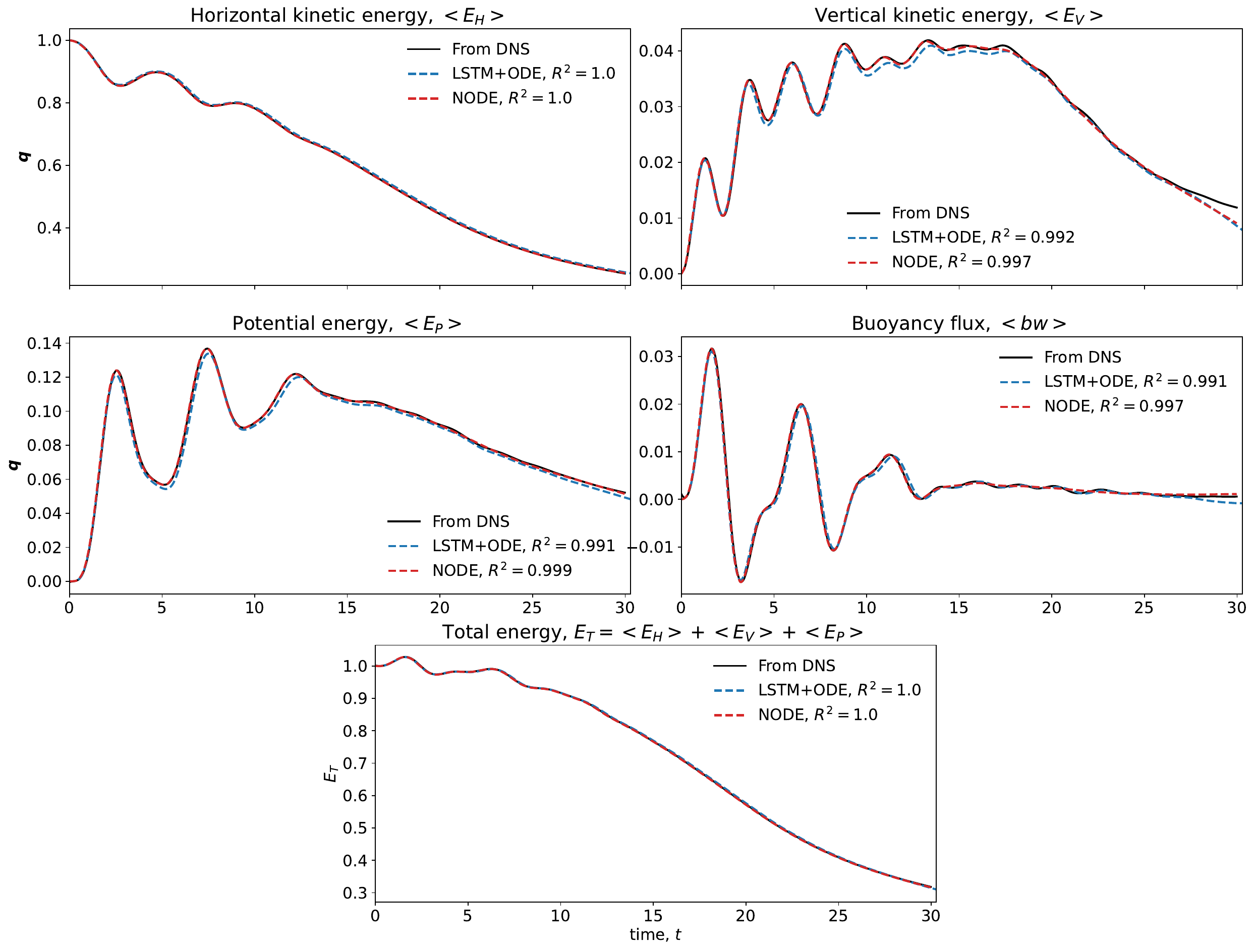}
\end{center}
 \caption{Results from \textit{a posteriori} tests using LSTM and NODE models. The overall accuracy of the models is measured using the coefficient of determination, $R^2$, of the ML results compared to true time evolution of each variable.
\label{fig:results-SST-aposteriori-best}}
\end{figure}

We show the results from the best LSTM and NODE models (based on the timescales analysis) in \fig\ref{fig:results-SST-aposteriori-best}. Both the LSTM and NODE models are able to accurately predict the time evolution of $\boldsymbol{q}$. In terms of the overall accuracy measured by the $R^2$ values, the NODE model offers better accuracy. Moreover, the predictions are numerically stable over time. Note that although the results show the model initialized from $t=0$, which is within the training regime, we also tested the model at various initial conditions within and outside the training regime. All the tests resulted in similarly accurate and numerically stable predictions. We wanted to demonstrate that the error acquired by the models when tested in the training regime (see the predictions for $E_P$ in the interval $t = [0,12]$) does not influence the predictions in the testing regime drastically. The time-series models are stable even with accrual of small errors over time, which has been shown to be detrimental for the numerical stability of models focused only on spatial information \cite{meena2023surrogate}.

The results demonstrate the capability of the time-series ML models to accurately predict the time evolution of complex nonlinear dynamical systems like the reduced-order SST. Similar to the pendulum results, the capability of the ML time-series models to accurately predict the total energy of the turbulent flow is notable (\fig\ref{fig:results-SST-aposteriori-best} bottom row), especially given that no information regarding the total energy was provided to the models. 
Note that, as a separate training campaign, we also trained the models with a physics-informed approach by imposing constraints of time rate of change of total energy, $\text{d}E_T/\text{d}t$, on the loss functions, thereby explicitly preventing the system from gaining energy. However, these results were not notably different from the current analysis. It is interesting to see the time-series models are able to predict the total energy of the system when the correct timescales are used for the inputs to the ML models.

\subsubsection{Interpreting the ML model data hyperparameter for SST}\label{subsubsec:SST-parametric}

We apply the physical timescale extraction framework discussed in \sectn\ref{subsubsec:pendulum-parametric} on the highly nonlinear SST problem. The current study is motivated by this turbulence modeling problem because there are no analytical solutions for the timescales in turbulent flow. In SST, there are, at a minimum, three important timescales: the buoyancy time scale, the inertial time scale, and the viscous time scale. Currently, the timescales can only be deduced from a combination of analytical work, laboratory or field measurements, and DNS -- extremely computationally expensive campaigns, e.g. \cite{debk19,riley2023effect,couchman2023mixing}. Even with these tools, the dominant timescales are hypothesized and then tested against data. This process becomes increasingly difficult when the flow becomes more complex due to, for example, mean shear. Of interest here is whether the relevant timescales could emerge from the ML modeling framework without any hypothesis. The pendulum results demonstrated this claim for a simple dynamical system.

\begin{figure}
\begin{center}
  \includegraphics[width=0.9\textwidth]{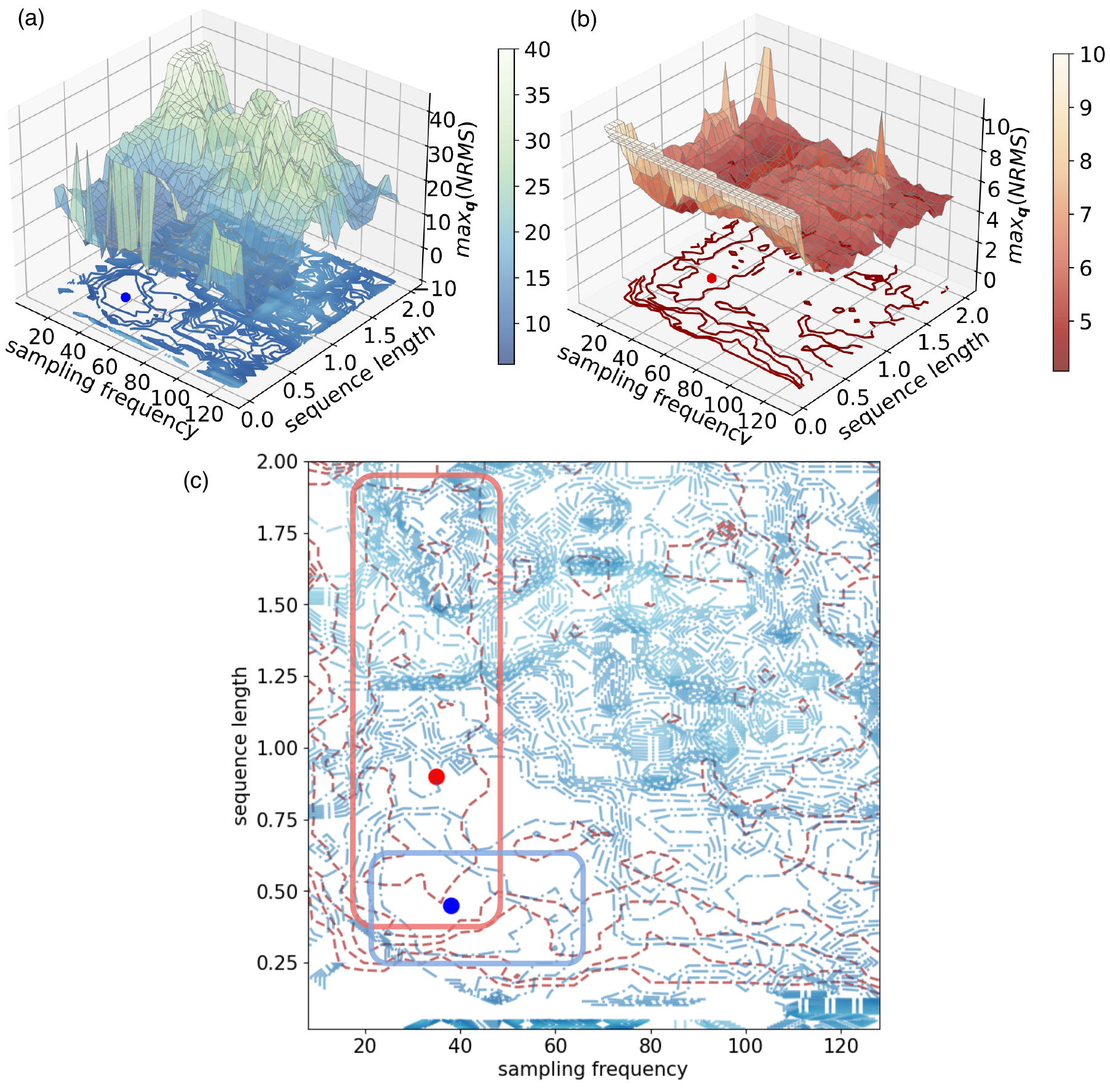}
\end{center}
 \caption{Effect of sequence length and sampling frequency of input data on (a) LSTM model and (b) NODE model. (c) Comparing the time-scales identified by LSTM (blue $-.$ contour) and NODE (red $--$ contours) models based on effect of sequence length and sampling frequency.
\label{fig:results-SST-parametric}}
\end{figure}

We conduct a similar ML timescales parametric study on the SST data using both the LSTM and NODE models, as shown in \fig \ref{fig:results-SST-parametric}. The parametric plot of timescales identifies unique regions of minimum error: between (1) sampling frequency $30$-$50$ and sequence length $0.4$-$0.6$ for LSTM, and (2) sampling frequency $20$-$40$ and sequence length $0.5$-$1$ for NODE. Interestingly, while the sampling frequency region is similar for both the models, the NODE model is accurate and numerically stable above a sequence length of $0.5$. This may pertain to the increased numerical stability and accuracy observed for NODE. These local regions correspond to the minimum information required by the inputs of the ML models to effectively predict the SST dynamics. Based on our inferred knowledge gained from studying the pendulum problem, we can potentially claim that these timescales are related to the physical timescales of the system. 

To validate the above claim, we identify the points of lowest error in these regions, denoted by the blue and red dots in \fig\ref{fig:results-SST-parametric} (a-b), respectively for the LSTM and NODE models. The ratio of the ML timescales, sequence length to sampling period ($1/$sampling frequency), at these points of lowest error are approximately $16$ for LSTM and $32$ for NODE. We note that these ML timescale ratios are integer multiples of the Reynolds number for the SST data, $Re=3200$. Physically, the Reynolds number of the SST flow can be defined by the ratio of the fluctuating timescale due to the buoyancy flux and decay timescale due to viscous dissipation \cite{riley2003dynamics}. Broadly, they represent the smaller and larger timescales of the system, which are what the time-series ML models extract as well. 
Thus, the ratio of the ML timescales are integer multiples of the physical timescale ratio known for SST flows. 
These observations validate our claim that our ML timescales analysis is able to interpret the behavior of the ML models and extract physical timescales of even complex turbulent fluid flow systems.

\begin{figure}
\begin{center}
  \includegraphics[width=1\textwidth]{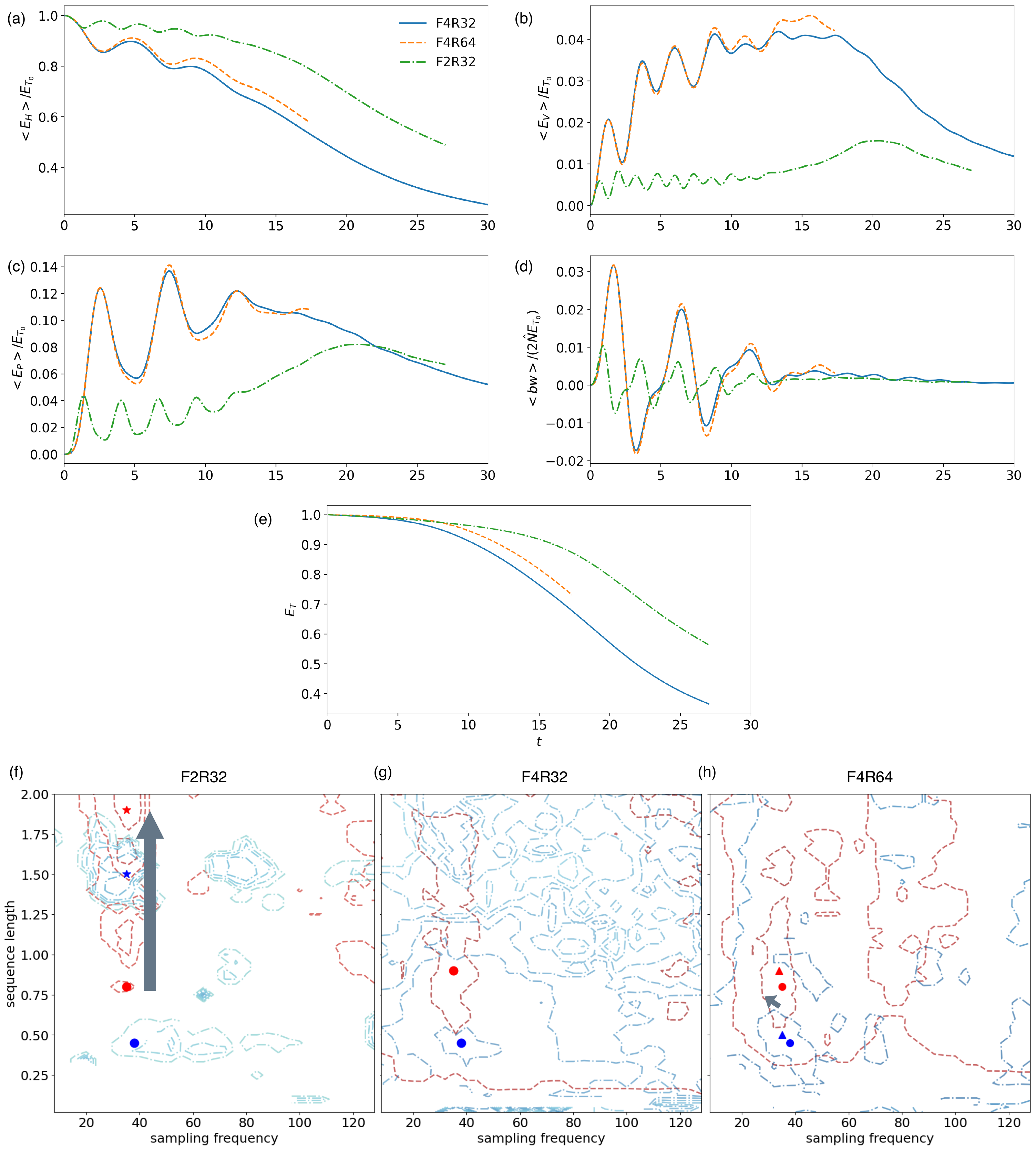}
\end{center}
 \caption{(a-e) The time series of SST data at different $Fr$ and $Re$ number values. (f-h) The parametric study for the LSTM (blue $-.$ contour) and NODE (red $--$ contour) models conducted independently on the 3 different dataset. The minimum information points are indicated for each case: $\bigstar$ for $Fr=2$, $Re=3200$;  $\bigcirc$ for $Fr=4$, $Re=3200$ (also indicated in the other two cases for comparison); and  $\bigtriangleup$ for $Fr=4$, $Re=6400$.
\label{fig:results-SST-parametric-FRstudy}}
\end{figure}

We generalize the above timescales extraction framework for various parameter ranges of SST flows initialized by Taylor--Green vortices \cite{hebert06b}. These flow are parameterized by the $Fr$ and $Re$. We chose to perform the timescale analysis on two flows: (1) $Fr=2$, $Re=3200$ and (2) $Fr=4$, $Re=6400$, thereby testing the timescale extraction framework for change in $Fr$ and $Re$. The time series of the flow states for the three SST flows are shown in \figs\ref{fig:results-SST-parametric-FRstudy} (a-e). Our objective is to compare the timescales extracted using the original/baseline analysis with the dataset $Fr=4$ and $Re=3200$ (blue lines in \figs\ref{fig:results-SST-parametric-FRstudy} (a-d)) with that when (1) the $Re$ is increased (orange lines in \figs\ref{fig:results-SST-parametric-FRstudy} (a-d)) and (2) the $Fr$ is decreased (green lines in \figs\ref{fig:results-SST-parametric-FRstudy} (a-d)).

We do not change the model hyperparameters of the LSTM and NODE models, and obtain two different models by training the two datasets independently. The focus of generalizability here is not to have a single model to solve parameter ranges of the flow, but to investigate the generalizability of our timescale extraction framework. The results of the timescale analyses are shown in \figs\ref{fig:results-SST-parametric-FRstudy} (f-h). When the $Re$ is doubled and $Fr$ is kept constant, there is marginal change in the decay rate and buoyancy flux as seen from the time-series in \figs\ref{fig:results-SST-parametric-FRstudy}~(a-d). The timescales extracted by the ML models also reflect this, depicted by the slight changes in the sampling frequency and sequence length as seen from \fig\ref{fig:results-SST-parametric-FRstudy}~(h). When the $Fr$ is halved and $Re$ is unchanged, the flow is characterized by longer times to decay and thus, lower decay rate, as shown in \figs\ref{fig:results-SST-parametric-FRstudy}~(a-c) and (e). Thus, one could expect the time-series ML models to require more time lengths of data to learn the dynamics -- i.e., longer sequence length. This is what is reflected in the timescales extracted by the ML models as shown in \fig\ref{fig:results-SST-parametric-FRstudy}~(f). While the sampling frequency remains same, the sequence length increases, denoting the capability of the framework to capture the change in the decay dynamics of the system, thus, demonstrating the generalizability of the timescales extraction framework for various parameter regimes of SST flows.


\section{Concluding remarks}
\label{sec:conclusion}

We present a time-series based machine learning (ML) modeling framework for second moment closure modeling of Unsteady Reynolds Averaged Navier Stokes (URANS) simulations of stably stratified turbulence (SST). Instead of individually formulating higher-order closures for the nonlinear terms in the URANS equations, we train time-series ML models using high fidelity direct numerical simulation (DNS) data to model the entire right-hand-side (RHS) of the URANS equations. The current approach is applied to one-dimensional SST data, which is spatially reduced by taking advantage of the homogeneity of the flow in three-dimensions. We choose two ML architectures for modeling: (1) Long short-term memory (LSTM) and (2) Neural Ordinary Differential Equation (NODE). The interpretability of the data requirements for the models are assessed by testing the accuracy of the models against two \textit{data hyperparameters}, the sampling frequency and sequence length of the input data to the models, revealing two ML timescales of the data. The accuracy of the models based on these hyperparameters corresponds to the minimal information required by the ML models to effectively capture the dynamics of the complex system. We find that variations/trends in the ML timescales of the best model corresponds to variations/trends in physical timescales known for such flows.

The framework is first demonstrated for a simple dynamical system -- a damped pendulum. The time-series ML models accurately capture the dynamics of the pendulum. Moreover, the ratio of the ML timescales extracted based on the sampling frequency and sequence length of the best models correspond to the ratio of the oscillating frequency to the damping frequency of the pendulum. Physically, this finding denotes that the minimum information required by the time-series ML models to accurately capture the dynamics of a system also captures fundamental timescales of the system. 

We extend the demonstration to a decaying SST flow initialized using Taylor--Green vortices. The time series data of the SST involves laminar, transition, and turbulent decay, posing a much more complex nonlinear problem for the time-series ML modeling framework. The LSTM and NODE models are not only accurate for predicting the time evolution of the SST, but also are numerically stable. Moreover, the ratio of the timescales extracted by the ML models is related to the Reynolds number of the flow. We also generalize the timescale extraction framework for various SST, with different Froude and Reynolds numbers. The ML timescales extracted are validated with the relative change in dynamics of the SST according to Froude and Reynolds numbers.

There are a couple of avenues to extend the current framework. First, while the LSTM model has shown great capability to model the dynamics of the SST, there are better time-series models in literature, e.g. GRU \cite{chung2014empirical,weerakody2021review}, which have been shown to be more accurate than LSTM. The current work is a preliminary effort to demonstrate that such time-series models can indeed be used to formulate interpretable models for SST dynamics. There is a clear avenue to extend the analysis for more complex models. Second, the current analysis is not generalized for predicting various parameter ranges of flow using a single model. This requires further investigation and potentially, a framework to incorporate the variation in the nature of the dynamics. Finally, while the prediction of the models are deterministic, the behavior of SST in nature warrants uncertainty quantification of the results of the models. The unique opportunity provided by GPU enabled high-performance computing platforms to perform ML inference at scale can further enable probabilistic approaches for modeling. Such complimentary advancements to the current framework can enable modeling of flows with higher complexity and spatial dimensions.



\section*{Acknowledgements}
This research was supported by the Office of Naval Research via grant N00014-19-1-2152. High performance computing resources were provided via the U.S.\ Department of Energy(DOE) INCITE program by the Oak Ridge Leadership Computing Facility, which is a DOE Office of Science User Facility supported under Contract DE-AC05-00OR22725. Computing resources were also provided through the U.S.\ Department of Defense High Performance Computing Modernization Program. 

\section*{Data and code availability}
Correspondence and requests for material should be addressed to M.G.M. 
Our algorithms will be made freely available in \href{https://github.com/muralikrishnangm/sst-rans-ml.git}{\color{blue}Link} after peer-review.

\section*{Author contributions}
M.G.M, S.M.deB.K, and J.J.R initiated and designed the project. S.M.deB.K curated the stably stratified turbulence direct numerical simulation data. M.G.M., D.L., and A.D.S. designed and performed the machine learning analysis with feedback from A.K. and W.H.B.. M.G.M., D.L., A.D.S., and S.M.deB.K wrote the manuscript with feedback from all the coauthors.

\section*{Author declaration}
The authors declare no competing interests.





\bibliographystyle{unsrtnat}
\bibliography{refs}


\end{document}